\documentclass[12pt,preprint]{aastex}

\newcommand{\antares}{ANTARES }

\shorttitle{ANTARES point source search}
\shortauthors{Adri\'an-Mart\'inez et al.}

\begin{document}

\title{First Search for Point Sources of High Energy Cosmic Neutrinos with the ANTARES Neutrino Telescope}


\author{S.~Adri\'an-Mart\'inez\altaffilmark{1}, 
J.A.~Aguilar\altaffilmark{2}, 
I.~Al Samarai\altaffilmark{3}, 
A.~Albert\altaffilmark{4}, 
M.~Andr\'e\altaffilmark{5}, 
M.~Anghinolfi\altaffilmark{6}, 
G.~Anton\altaffilmark{7}, 
S.~Anvar\altaffilmark{8}, 
M.~Ardid\altaffilmark{1}, 
A.C.~Assis Jesus\altaffilmark{9}
% ...
}
 \altaffiltext{1}{Institut d'Investigaci\'o per a la Gesti\'o Integrada de les Zones Costaneres (IGIC) - Universitat Polit\`ecnica de Val\`encia. C/  Paranimf 1 , 46730 Gandia, Spain.}
\altaffiltext{2}{IFIC - Instituto de F\'isica Corpuscular, Edificios Investigaci\'on de Paterna, CSIC - Universitat de Val\`encia, Apdo. de Correos 22085, 46071 Valencia, Spain}
\altaffiltext{3}{CPPM, Aix-Marseille Universit\'e, CNRS/IN2P3, Marseille, France}
\altaffiltext{4}{GRPHE - Institut universitaire de technologie de Colmar, 34 rue du Grillenbreit BP 50568 - 68008 Colmar, France }
\altaffiltext{5}{Technical University of Catalonia, Laboratory of Applied Bioacoustics, Rambla Exposici\'o, 08800 Vilanova i la Geltr\'u, Barcelona, Spain}
\altaffiltext{6}{INFN - Sezione di Genova, Via Dodecaneso 33, 16146 Genova, Italy}
\altaffiltext{7}{Friedrich-Alexander-Universit\"at Erlangen-N\"urnberg, Erlangen Centre for Astroparticle Physics, Erwin-Rommel-Str. 1, 91058 Erlangen, Germany}
\altaffiltext{8}{Direction des Sciences de la Mati\`ere - Institut de recherche sur les lois fondamentales de l'Univers - Service d'Electronique des D\'etecteurs et d'Informatique, CEA Saclay, 91191 Gif-sur-Yvette Cedex, France}
\altaffiltext{9}{Nikhef, Science Park,  Amsterdam, The Netherlands}
% ...

\begin{abstract}
 Results are presented of a search for cosmic sources of high energy 
 neutrinos with the \antares neutrino telescope. The data were collected 
 during 2007 and 2008 using detector configurations containing 
 between 5 and 12 detection lines. The integrated live time of the 
 analyzed data is 304 days. Muon tracks are reconstructed using 
 a likelihood-based algorithm. Studies of the detector timing indicate 
 a median angular resolution of $0.5 \pm 0.1$ degrees. The neutrino flux 
 sensitivity is $7.5 \times 10^{-8} ~{ (E_{\nu}/ \rm GeV)^{-2}~ GeV^{-1} ~s^{-1} ~cm^{-2}}$ 
 for the part of the sky that is always visible ($\delta < -48$ degrees), 
 which is better than limits obtained by previous experiments. 
 No cosmic neutrino sources have been observed.
\end{abstract}


\hspace{1.5 cm}
\par{\noindent \it We dedicate this paper to the memory of our colleague and friend Luciano Moscoso,
who passed away during the preparation of this paper.}

\section{Introduction}
  
 High energy cosmic rays permeate the universe, but their origins remain uncertain. 
 Many types of acceleration sites have been suggested, such as supernova remnants, 
 micro-quasars and active galactic nuclei (see e.g. \citet{Becker2008173} for a review).
 If such objects are sources of cosmic rays, high energy neutrinos may also 
 be produced. 
 These may be detected as a diffuse flux \citep{diffuse}, but the observation 
 of point-like sources of cosmic neutrinos would offer a unique opportunity 
 to identify and study the sites and mechanisms of cosmic ray acceleration. 
  
 In this paper, results are presented of a search for cosmic point-like 
 sources of high energy muon neutrinos performed with the first data 
 taken by the \antares telescope. 
 Located in the Mediterranean Sea, the \antares experiment
 is predominantly sensitive to neutrinos from the southern hemisphere
 in the TeV to PeV energy range. In particular, this allows the study of
 Galactic sources and complements the IceCube observatory, which primarily 
 explores the northern hemisphere.

\section{Data Collection}

 The detection principle relies on measuring Cherenkov light emitted
 by high energy 
 charged particles that result from neutrino interactions inside or near 
 the instrumented volume. In particular, charged current interactions of 
 muon-neutrinos produce high energy muons. The muon direction closely 
 follows that of the neutrino and can often be reconstructed with sub-degree accuracy. 
 The large background from downgoing muons due to cosmic ray interactions 
 in the atmosphere is reduced by selecting only upgoing muons as 
 neutrino candidates.

 The Cherenkov light is detected by an array of photomultiplier tubes, each 
 housed in a pressure resistant glass sphere called an Optical Module (OM) 
 \citep{Amram:2001mi}. The OMs are placed in storeys of triplets along vertical
 detector lines, which are anchored to the sea bed at a depth of 2475 meters 
 with a spacing of approximately 60 meters. Each of the 12 detector lines 
 contains up to 25 storeys, separated by 14.5 meters.

 The arrival time and charge of the photomultiplier signals are 
 digitized \citep{arspaper} into `hits' and transmitted to shore.
 An online filter identifies events containing muons within the data stream 
 that otherwise consists of optical backgrounds due to natural radioactive 
 decays and bioluminescence. The selected events are stored for offline 
 reconstruction. A detailed description of the detector, the data acquisition 
 and the online filtering algorithms is given in \citet{detector, Aguilar:2006pd}.

\subsection{Data Taking and Processing}

 The data presented here were collected between January 31st, 2007 and  
 December 30th 2008. 
 During this time, the construction of the \antares detector was still in progress.
 The detector consisted of 5 lines for most of 2007 and of 9, 10 and 12 
 detector lines during 2008. The total live time of the data used for 
 the analysis is 304 days.( 144, 38, 48 and 74 days with
 5, 9, 10 and 12 detector lines respectively).

 The online event selection identifies triplets of OMs that detect 
 multiple photons, either as a high-charge hit, or as hits separated 
 by less than 20 ns on adjacent OMs.
 At least 4 of such triplets are required throughout the detector, with 
 the relative photon arrival times being compatible with the hypothesis of
 the light being emitted along the track of a relativistic particle.


 The arrival times of the hits are calibrated as described in \citet{timing_paper}. 
 The inter-line timing has been measured in an iterative procedure by comparing the 
 expected hit times to those measured in a large sample of reconstructed downgoing 
 muons. In addition, a similar, independent measurement has been done using light 
 from the optical beacon system, with compatible results.
 The positions and orientations of the OMs vary because of the sea currents.
 An acoustic positioning system, combined with compasses and tiltmeters located 
 along the detector lines, measures the positions and orientations of the OMs, 
 with an accuracy of $\sim$10 cm. 

\subsection{Event Reconstruction and Selection }

 From the timing and position information of the hits, muon
 tracks are reconstructed using a multi-stage fitting
 procedure, based on \citet{thesis}. 
 The initial fitting stages provide the hit selection and 
 starting point for the final fit.
 The final stage consists of a maximum likelihood fit of the 
 observed hit times and includes the contribution of optical 
 background hits. 
 In order to maximize the robustness against inaccuracies 
 in the simulation of the optical background, the charge 
 information of the hits is not used in the present analysis.

\begin{figure}[!htp]
\begin{center} 
  \includegraphics[width=0.8\textwidth]{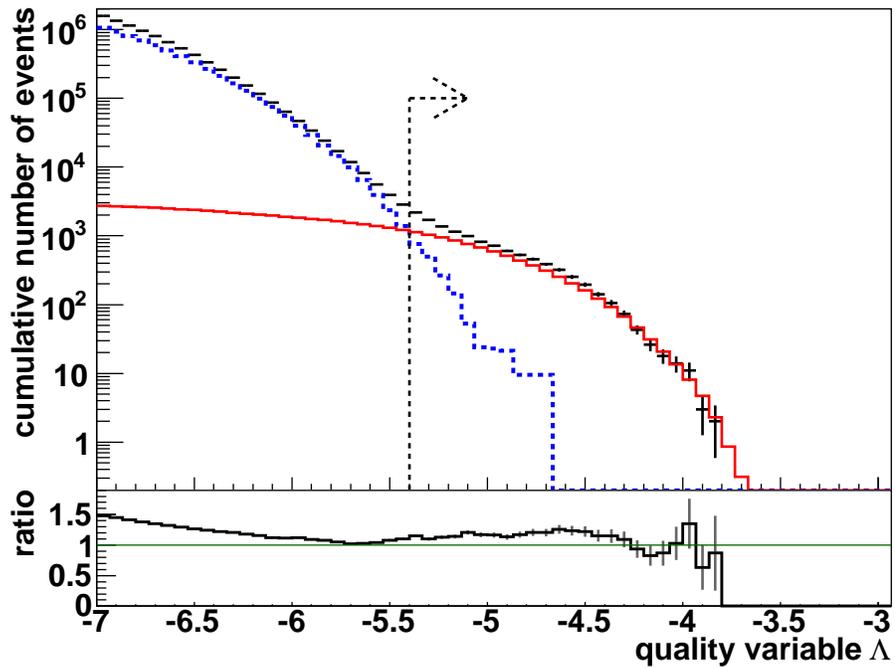}
\caption{Cumulative distribution of the reconstruction quality variable,
         $\Lambda$, for upgoing events in data (black points) which have
         an error estimate $<1$ degree.
         The Monte Carlo simulated contributions from mis-reconstructed atmospheric 
         muons (dashed blue line) and atmospheric neutrinos (solid red line) are 
         also shown. 
         The bottom panel shows the ratio between data and simulation.
         The vertical line indicates the analysis cut $\Lambda>-5.4$.}  
\label{fig:lambda}
\end{center} 
\end{figure}

 Neutrino candidates are selected from upgoing events using 
 criteria that have been determined in a 'blind' manner; i.e. 
 before performing the search analysis on the data. The criteria are chosen
 to optimize the sensitivity, which is defined as the expected median 
 value of the upper limit on the neutrino flux.      
 The angular uncertainty obtained from the muon track fit is 
 required to be smaller than 1 degree. Tracks are also required to
 have a good reconstruction quality. The latter is quantified by a parameter 
 $\Lambda$, which is based on the value of the likelihood function 
 obtained for the fitted muon (see \citet{thesis} for details). 
 The cumulative distribution of $\Lambda$ 
 for muons which are reconstructed as upgoing is 
 shown in Figure \ref{fig:lambda} along with the simulated contributions 
 from atmospheric muons and neutrinos. The simulation uses the `Bartol' model 
 \citep{PhysRevD.53.1314} for the atmospheric neutrino flux. 
 The atmospheric muons are simulated by the QGSJET \citep{Kalmykov:1993qe} and 
 CORSIKA \citep{corsika} packages with the primary cosmic ray flux from \citet{Nikolsky:1985cp}.

 The final sample of neutrino candidates consists of 2190 upgoing      
 events with $\Lambda>-5.4$ and an angular uncertainty $<1$ degree, which were
 selected out of a total of $\sim$ 100 million (down going) 
 reconstructed muons. The simulation indicates that the selected 
 sample contains about 60\% atmospheric neutrinos;  the rest are 
 mis-reconstructed atmospheric muons. The agreement between data 
 and simulation in the selected sample is well within the $\sim$30\% 
 systematic uncertainty on the neutrino flux \citep{fluxunc}.

%
%


\section{Detector Performance}

 The angular resolution and effective area of the detector for neutrinos 
 passing the selection cuts have been determined using simulations and
 are shown in Figure \ref{fig_angres}. The simulations include the 
 stochastic energy loss of muons via electromagnetic shower production, 
 the propagation of Cherenkov photons through sea water and a detailed 
 simulation of the detector electronics.

\subsection{Angular Resolution}

 The cumulative distribution of the angle between the 
 reconstructed muon direction and the neutrino direction is shown 
 in Figure \ref{fig_angres} (left panel) for neutrino events with a 
 neutrino spectrum proportional to $E_{\nu}^{-2}$, where $E_{\nu}$ 
 is the neutrino energy. The median of this angular error is 0.5 $\pm$ 0.1 
 degrees. For the subset of data in which the full 12-line detector 
 was operational, the resolution is estimated to be $0.4 \pm 0.1$ degrees.
 



 The systematic uncertainty on this quantity has been estimated by 
 varying the time resolution of the OMs $\Delta_{t}$ in the simulation. 
 The allowed range of $\Delta_{t}$ is determined by requiring that 
 the $\Lambda$ distribution in the resulting simulation be compatible 
 with the observed atmospheric neutrino events. 
 The best agreement between data and simulation is obtained for 
 $\Delta_{t}=2.5$ ns. 
 Hence, this value is used for all simulations in this analysis, in 
 particular for extracting the central value of the allowed range of angular 
 resolutions. A time resolution of 3.4 ns is found to be incompatible with 
 the neutrino flux model at the 2$\sigma$ level, where the uncertainty is
 taken from \citet{fluxunc}. This places an upper bound on the 
 time resolution, which translates into a 1 $\sigma$ systematic 
 uncertainty on angular resolution of 0.1 degrees.
 This uncertainty incorporates, to first order, all effects which have 
 a net result of degrading the time resolution, such as possible 
 mis-alignments and inaccuracies in the simulation of light propagation 
 in the water or the transit time distribution of the PMT.
 A similar analysis with analogous results has been performed using downgoing 
 muon data instead of upgoing neutrino candidates.

 The absolute orientation of the detector is known with an accuracy
 of about $0.1$ degrees \citep{garothesis}; this uncertainty is taken 
 into account as an independent effect.

\subsection{Acceptance} 

\begin{figure}[!htp]
\begin{center} 
  \includegraphics[width=0.95\textwidth]{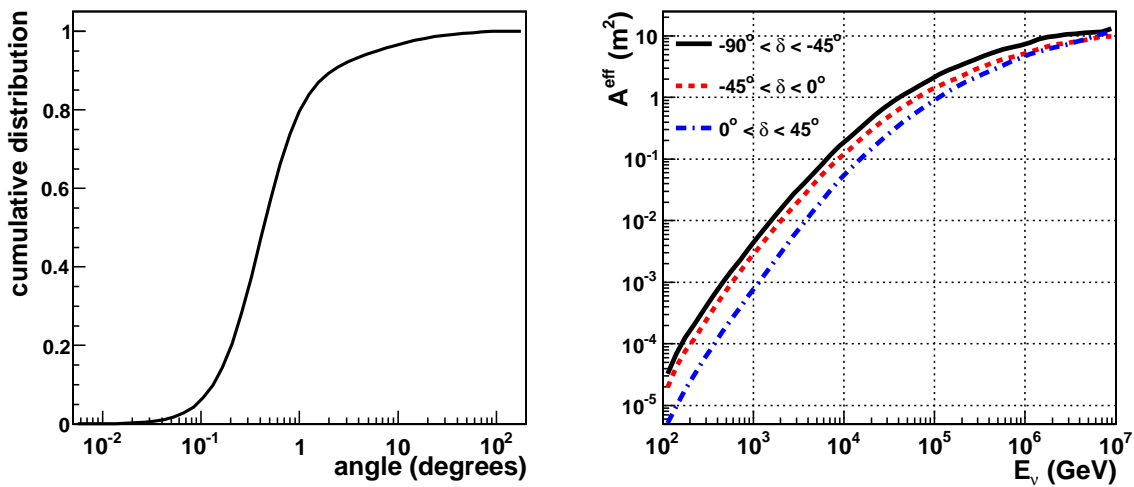}
\caption{{Performance of the detector for the selected signal events as
         determined from simulation. The cumulative distribution of the angle 
         between the reconstructed muon direction and the true neutrino direction
         is shown for selected signal events with an $E_{\nu}^{-2}$ spectrum (left panel). 
         The neutrino effective area $A^{\rm eff}_{\nu}$, averaged over three 
         declination ranges, is 
         shown as a function of the neutrino energy (right panel).}
}
\label{fig_angres}
\end{center} 
\end{figure}

 { The effective area for muon neutrinos $A^{\rm eff}_{\nu}$ is defined as the ratio 
 between the selected neutrino event rate and the cosmic neutrino flux. 
 It is determined from simulations and is shown in Figure \ref{fig_angres} 
 (right panel) as a function of the neutrino energy for three declination intervals. }
 { Throughout this paper, the cosmic neutrino
 flux is assumed to consist of an equal amount of $\nu_{\mu}$ and $\bar{\nu}_{\mu}$.}

 In the search, limits are set on the constant $\phi$ in
 the flux parametrization ${{dN}/{dE_{\nu}}} = \phi \times [E_{\nu}/{\rm GeV}]^{-2} \rm~ GeV^{-1} cm^{-2} s^{-1}$.
 The acceptance $A$ for such a flux is defined as the constant of 
 proportionality between the number of selected signal events and 
 the flux intensity $\phi$.  It can be computed by convoluting $A^{\rm eff}_{\nu}(E)$ 
 with ${{dN}/{dE_{\nu}}}(E)$. 
 For declinations $\delta<-48$ degrees, $A$ and $A^{\rm eff}_{\nu}$ 
 are approximately constant. For $-48 < \delta < 48$ degrees, the functions 
 decrease because of the requirement that the tracks are upgoing. 
 For a source declination of -90 (0) degrees,  $A$ = 3.2 (1.8) $\rm \times 10^7 ~GeV ~cm^{2} ~s$. 
 This means a total of 3.2(1.8) neutrinos would be detected 
 and selected from a point source with a flux of $10^{-7} \rm~ GeV^{-1} cm^{-2} s^{-1}$.
 For this flux model, the energy of 80\% of the selected signal 
 is in the range $3 < E_{\nu} < 700$ TeV.

 To constrain the systematic uncertainty on the acceptance, the
 atmospheric neutrino data have been compared to a simulation in 
 which the efficiency of each OM is reduced. The maximal reduction
 for which data and simulation are still compatible corresponds to 
 a 15\% reduction in the acceptance for signal neutrinos with an
 $E_{\nu}^{-2}$ spectral shape. Hence, a 15\% systematic uncertainty 
 on the acceptance is used in the limit calculations.


\section{Search Method}

 Two distinct approaches to look for point-like neutrino sources have been used.
 In the \emph{full sky search}, a search is made for an excess of events
 over the atmospheric neutrino background anywhere in the field of view. 
 In addition, a \emph{candidate list search} is made, where 
 the presence of a signal is tested at the locations
 of 24 known pre-defined high energy gamma ray sources which could 
 be expected to emit neutrinos. They include supernova
 remnants, microquasars and BL Lac objects. While more restrictive in scope, 
 the candidate list search requires less signal to reach a significant excess,
 compared to the all-sky search. The addition of the candidate search may 
 therefore be thought of as a way to give extra \emph{a-priori} weight to 
 the preselected sources.

\subsection{Event Likelihood}

 The search method is based on the likelihood of observing 
 the events, which is defined as:
\begin{equation}
\log {\cal L}_{\rm s+b} = \sum_i \log [ \mu_{\rm sig} 
                         \times {\cal F}(\beta_i(\delta_s,\alpha_s)) + {{\cal B}(\delta_i) } ] - \mu_{\rm sig} 
                          - N_{\rm bg}
\label{eq:lik1}
\end{equation} 
 where the sum is over the neutrino candidate events, and $\cal F$ is a 
 parametrization of
 the point spread function. This is defined as the probability density to
 find the reconstructed muon $i$ an angle $\beta$ away from the  
 declination $\delta_s$, and right ascension $\alpha_s$ of the source; it is closely
 related to the angular resolution (see Figure \ref{fig_angres}).
 ${\cal B}(\delta)$ is a smooth parametrization of
 the background rate derived from the observed declination distribution 
 of the 2190 selected events. The mean number of selected signal 
 events produced by the source is $\mu_{\rm sig}$. The term $N_{\rm bg}$ 
 represents the total number of expected background events, which is constant
 and therefore does not influence the maximum likelihood fits or the likelihood 
 ratio.

 In the candidate list search, the likelihood is maximized for each candidate
 by numerically fitting the source intensity $\mu_{\rm sig}$ to the events located
 within 20 degrees of the source, with the source coordinates fixed to the known
 position.
 In the full sky search, potentially significant clusters are first identified
 using a loose cone selection, which requires at least 4 events 
 in a cone of $3$ degrees diameter. For each cluster, the likelihood 
 is maximized by fitting the source coordinates and the intensity, 
 yielding maximum-likelihood estimates for these quantities.
 
 The next step is to compute the test statistic, which is defined as the 
 logarithm of the likelihood ratio:
\begin{equation}
Q = \log {\cal L}^{\rm max}_{\rm s+b} - \log {\cal L}_{\rm b},
\end{equation}
 where ${\cal L}^{\rm max}_{\rm s+b}$ is the maximum value of the likelihood
 found in the fit and ${\cal L}_{\rm b}$ is the likelihood computed for the 
 background-only hypothesis ($\mu_{\rm sig}=0$). A large (small) value of 
 $Q$ indicates that the data are compatible with the signal (background). 

\subsection{Statistical Interpretation}


 The statistical interpretation of the search result relies on simulated
 pseudo-experiments in which the background events are randomly 
 generated by sampling the declination from the parametrization
 ${\cal B}$ and the right ascension from a uniform distribution.
 Events from a neutrino point source are simulated by adding events around
 the desired coordinates according to the point spread function. 
 The systematic uncertainties
 on the angular resolution and orientation of the detector are 
 incorporated by varying the simulated characteristics of the signal 
 events within the assigned uncertainties. 

 Distributions of $Q$ are obtained from the simulated experiments 
 by applying the search method to each of them. Significances in 
 the form of $p$-values are derived from the 
 $Q$ distribution for background-only pseudo-experiments. The distributions
 including simulated signal events are used to set flux limits following
 the Feldman-Cousins prescription \citep{fc}. The uncertainty in 
 the acceptance $A$ is accounted for in the limits by numerically 
 integrating over the allowed range defined by the $\pm 15\%$ uncertainty.

\section{Results}
\label{sec:results}

\begin{figure}[!htp]
\begin{center} 
  \includegraphics[width=\textwidth]{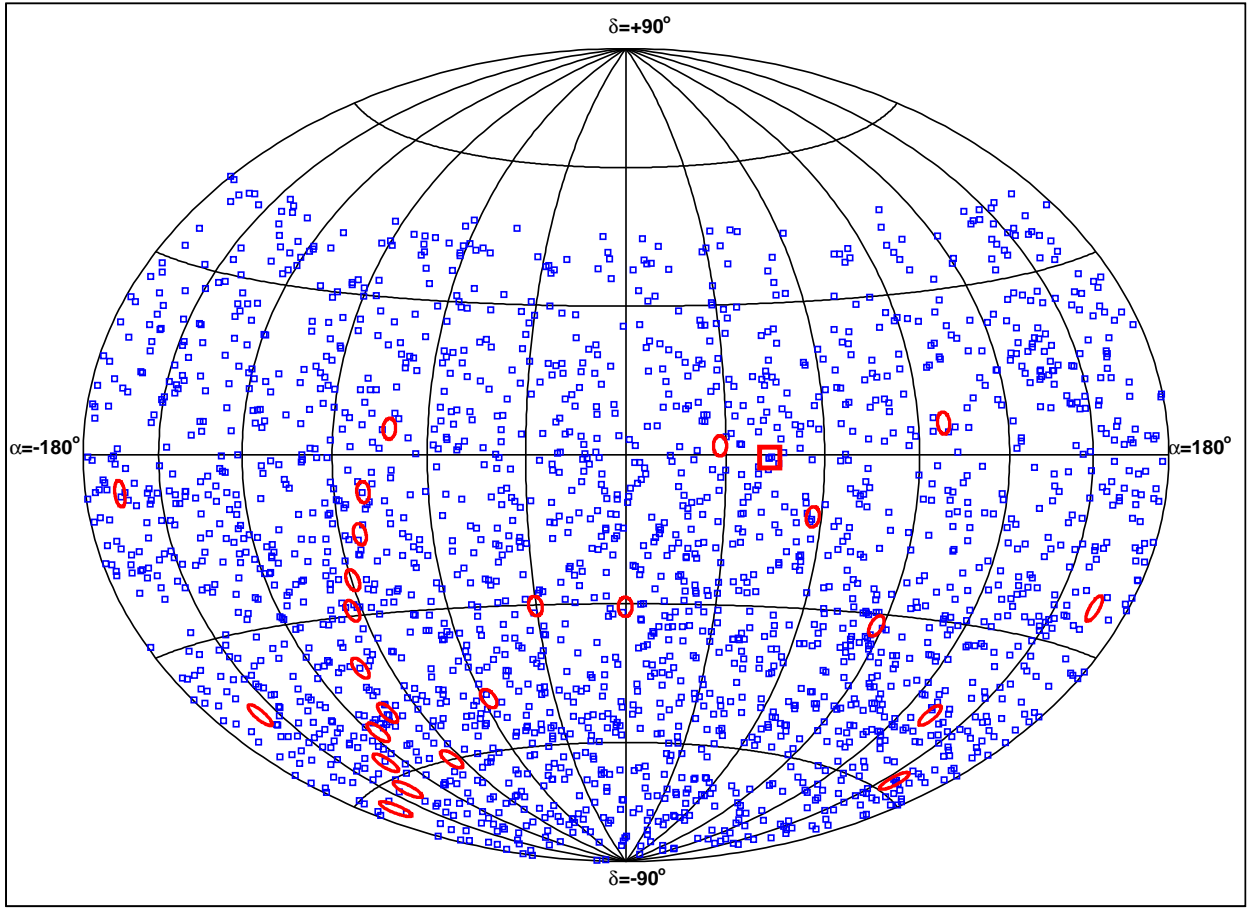}
\caption{Map in equatorial coordinates of the 2190 selected neutrino candidates. The
         position of the most significant cluster (see text) is 
         indicated by the square. The circles denote the positions of the 24
         sources from the candidate list. }
\label{fig_skymap}
\end{center} 
\end{figure}

 Figure \ref{fig_skymap} shows a sky map of the selected events 
 in equatorial coordinates along with the locations of sources from 
 the candidate list (circles) and of the most significant cluster found 
 in the full-sky search (square).


 In the full-sky search, no significant clusters of neutrino candidates have
 been found.
 The most signal-like cluster of events has a maximum likelihood estimate
 of the source position $(\alpha_s, \delta_s ) =  ( 43.21^{\circ}, -0.50^{\circ} )$. The fit 
 assigns 3.4 events as signal events and finds a value of the test statistic $Q$ =
 6.8. Such a value, or larger (more signal like), occurs with a probability 
 $p= 88\%$ anywhere in the field of view amongst background only pseudo-experiments. 
 The result is thus compatible with the background only hypothesis.

 The results of the candidate list search are shown in Table 
 \ref{tab:lims}. None of the candidate sources have a    
 significant excess of events at their location. 
 The most signal-like source candidate is HESS J1023-575, 
 where 3(5) events are within 1(3) degrees of its position.
 For this cluster of events, $Q=2.5$. In the absence of signal such a 
 value or larger, has a probability of $p=17\%$ to occur amongst the 
 24 source candidates. 

\subsection{Neutrino Flux Upper Limits}

\begin{figure}[!htp]
\begin{center} 
  \includegraphics[width=0.8\textwidth]{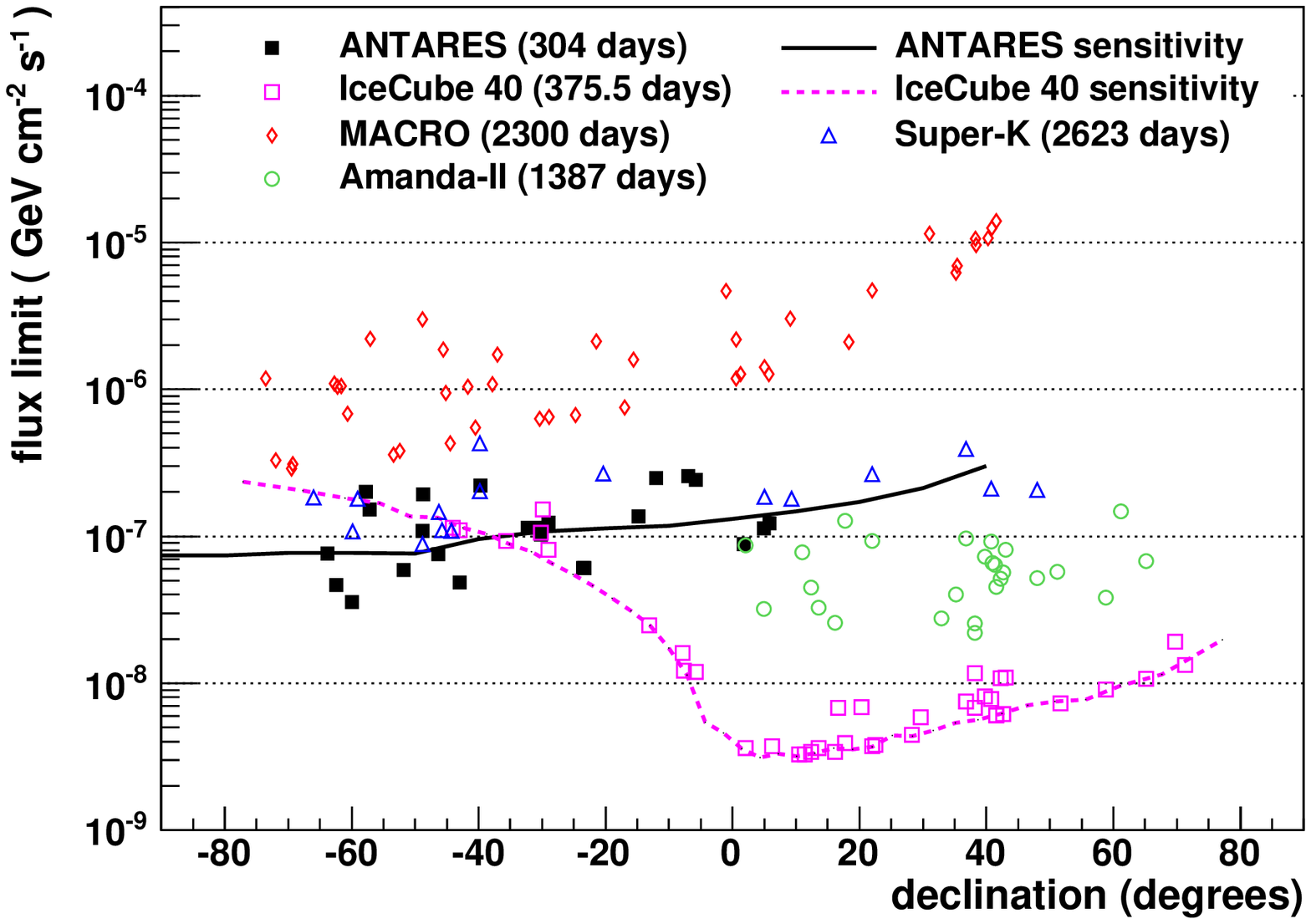}    
\caption{Limits set on the normalization $\phi$ of an $E^{-2}_{\nu}$ spectrum of high energy 
         neutrinos from selected candidates (see Table \ref{tab:lims}). The points show               
         the 90\% C.L. limit at the declination of the candidate source. In addition to 
         the present result, several previously published limits on sources in both the 
         southern and northern hemisphere are shown from \citet{Ambrosio:2000yx, Thrane:2009tw, 
         Abbasi:2008ih} and \citet{ic40}. {Also shown are sensitivities (solid and
         dashed lines) for the current analysis and for the search from \citet{ic40} (see
         \citet{Abbasi:2008ih, Thrane:2009tw} for the sensitivities corresponding to those
         searches).}
      }
\label{fig:lims}
\end{center} 
\end{figure}

 As no significant point sources are observed, 90\% confidence level limits 
 are obtained for the intensity, $\phi^{90\% \rm CL}$, of an $E_{\nu}^{-2}$ neutrino  
 flux from each of the source candidates. They are listed in 
 Table 1 and are shown in Figure \ref{fig:lims} as a function of the 
 source declination. Figure \ref{fig:lims} also shows the sensitivity of this
 analysis, defined as the median value of the expected limit. 
 It is in agreement with the median value of the actually observed limits.
 For the area of the sky that is always visible, the ($\delta < -48$ degrees) 
 sensitivity is about $7.5 \times 10^{-8} ~{ (E_{\nu} \rm/GeV)^{-2}~ GeV^{-1} ~s^{-1} ~cm^{-2}}$. 
 
\begin{table}
\begin{tabular}{lr@{.}lr@{.}l@{\hspace{0.2cm}}llll}
\hline
         source 
         & \multicolumn{2}{c}{$\alpha_s$ (deg)} 
         & \multicolumn{2}{c}{$\delta_s$ (deg)}
         & \multicolumn{1}{c}{$\mu_{\rm sig}^{\rm fit}$}
         & \multicolumn{1}{c}{$Q$}
         & \multicolumn{1}{c}{$p$-value}
         & \multicolumn{1}{c}{$\phi^{90\% \rm CL}$} \\
\hline
 HESS\ J1023-575 &   155&83 & -57&76 &  2.7 &  2.5 &   0.17 &        2.0\\
         GX\ 339 &   -104&30 & -48&79 &  2.2 &  2.1 &   0.26 &        1.9\\
RX\ J1713.7-3946 &  -101&75 & -39&75 &  1.0 &  1.7 &   0.46 &        2.2\\
 HESS\ J1837-069 &   -80&59 &  -6&95 &  1.1 &  1.5 &   0.55 &        2.6\\
   1ES\ 0347-121 &    57&35 & -11&99 &  1.4 &  1.1 &   0.70 &        2.5\\
         3C\ 279 &  -165&95 &  -5&79 &  0.91 &  0.74 &   0.83 &        2.4\\
        Cir\ X-1 &  -129&83 & -57&17 &  0.82 &  0.65 &   0.85 &        1.5\\
   PKS\ 2005-489 &   -57&63 & -48&82 &  $ 0$ & $ 0$ &  $ 1 $ &     1.1\\
Galactic\ Center &   -93&58 & -29&01 &  $ 0$ & $ 0$ &  $ 1 $ &     1.2\\
        LS\ 5039 &   -83&44 & -14&83 &  $ 0$ & $ 0$ &  $ 1 $ &     1.4\\
     H\ 2356-309 &    -0&22 & -30&63 &  $ 0$ & $ 0$ &  $ 1 $ &     1.1\\
RX\ J0852.0-4622 &    133&00 & -46&37 &  $ 0$ & $ 0$ &  $ 1 $ &     0.76\\
   PKS\ 0548-322 &    87&67 & -32&27 &  $ 0$ & $ 0$ &  $ 1 $ &     1.1\\
   PSR\ B1259-63 &   -164&30 & -63&83 &  $ 0$ & $ 0$ &  $ 1 $ &     0.76\\
   PKS\ 2155-304 &   -30&28 & -30&22 &  $ 0$ & $ 0$ &  $ 1 $ &     1.0\\
 HESS\ J1614-518 &  -116&42 & -51&82 &  $ 0$ & $ 0$ &  $ 1 $ &     0.59\\
         SS\ 433 &   -72&04 &   4&98 &  $ 0$ & $ 0$ &  $ 1 $ &     1.1\\
 HESS\ J0632+057 &    98&24 &   5&81 &  $ 0$ & $ 0$ &  $ 1 $ &     1.2\\
         RCW\ 86 &  -139&32 & -62&48 &  $ 0$ & $ 0$ &  $ 1 $ &     0.47\\
  RGB\ J0152+017 &    28&17 &   1&79 &  $ 0$ & $ 0$ &  $ 1 $ &     0.89\\
    Centaurus\ A &  -158&64 & -43&02 &  $ 0$ & $ 0$ &  $ 1 $ &     0.49\\
    ESO\ 139-G12 &   -95&59 & -59&94 &  $ 0$ & $ 0$ &  $ 1 $ &     0.36\\
            W28 &   -89&57 & -23&34 &   $ 0$ & $ 0$ &  $ 1 $ &     0.61\\
   1ES\ 1101-232 &   165&91 & -23&49 & $ 0$ & $ 0$ & $ 1 $ &       0.61\\
\hline
\end{tabular}
\caption{Results of the candidate list search. The source coordinates and the 
         $Q$ and $p$-values are shown as well as the limits on the flux intensity 
         $\phi^{90\% \rm CL}$; the latter has units $10^{-7} \rm GeV^{-1} cm^{-2} s^{-1}$.
         The rows are sorted in order of increasing p-value.}
\label{tab:lims}
\end{table}

 The present limits are more stringent than those obtained for the 
 northern hemisphere by previous multi year experiments (also indicated 
 in the figure) and are competitive with those set by the IceCube 
 collaboration \citep{ic40} for $\delta < -30^{\circ}$.                
 It should be noted that even though they all set limits 
 on the intensity of an assumed $E_{\nu}^{-2}$ spectrum, the 
 experiments are sensitive in different energy ranges.
 For this spectrum, \antares detects most events at energies in a broad 
 range around 10 TeV, which is the relevant energy range for several 
 galactic sources \citep{Crocker:2004nk}. Northern hemisphere limits 
 shown from the IceCube experiment probe the neutrino flux predominantly 
 in the region above 1 PeV \citep{ic40}.
 
 The event selection and the search method have been cross-checked 
 with an independent analysis using the same selection 
 criteria and a search method based on the expectation-maximization 
 algorithm \citep{EM,juanjopaper}. In this method, the angular spread   
 of the signal events is a free parameter in the likelihood and the maximization is  
 performed analytically. 
 The results of both the full sky and the candidate list search are 
 consistent with the results discussed earlier.



\section{Autocorrelation Analysis}

 In a separate analysis, the cumulative distribution of the number of event pairs 
 as a function of their angular separation has been studied using the neutrino 
 candidate events. This autocorrelation analysis has been employed to search for 
 features in the data such as an excess of tracks at an arbitrary level of angular 
 separation.  The study is independent of detector simulations and neutrino source 
 models.


 A reference autocorrelation distribution is determined by scrambling 
 the data itself approximately one million times and averaging the resulting
 distributions. The comparison between the data and the reference 
 distribution is performed following Li and Ma (1983).


 The maximum excess above the reference distribution 
 has a significance of $1.1~\sigma$ at angular scales smaller than 7~degrees. Such 
 a deviation is expected from a collection of random background events with a 
 (trial factor corrected) probability of $55\%$.


\section{Conclusions}

 A search for cosmic sources of high energy neutrinos using the
 data taken with the \antares neutrino telescope during the 
 first two years of operation has been presented. 
 For the first half of the data considered, the detector was 
 about half its final size. An estimate of the angular resolution
 constrained by data yields 0.5 $\pm$ 0.1 degrees (0.4 degrees 
 for the 12-line detector), confirming the expectation that 
 excellent angular resolution is achievable using sea water as 
 the detection medium.
 Neither the full sky search nor the candidate list search show
 a significant excess of events. Limits have been obtained on 
 the high energy neutrino flux for a number of selected source 
 candidates. For many candidate sources, the limits presented 
 here are the most stringent to date.

\acknowledgments

The authors acknowledge the financial support of the funding agencies:
Centre National de la Recherche Scientifique (CNRS), Commissariat
\'a l'\'ene\-rgie atomique et aux \'energies alternatives  (CEA), Agence
National de la Recherche (ANR), Commission Europ\'eenne (FEDER fund
and Marie Curie Program), R\'egion Alsace (contrat CPER), R\'egion
Provence-Alpes-C\^ote d'Azur, D\'e\-par\-tement du Var and Ville de
La Seyne-sur-Mer, France; Bundesministerium f\"ur Bildung und Forschung
(BMBF), Germany; Istituto Nazionale di Fisica Nucleare (INFN), Italy;
Stichting voor Fundamenteel Onderzoek der Materie (FOM), Nederlandse
organisatie voor Wetenschappelijk Onderzoek (NWO), the Netherlands;
Council of the President of the Russian Federation for young scientists
and leading scientific schools supporting grants, Russia; National
Authority for Scientific Research (ANCS), Romania; Ministerio de Ciencia
e Innovaci\'on (MICINN), Prometeo of Generalitat Valenciana and MultiDark,
Spain. We also acknowledge the technical support of Ifremer, AIM and
Foselev Marine for the sea operation and the CC-IN2P3 for the computing facilities.

\end{document}